# High-Pressure X-Ray Diffraction Study of Scheelite-type Perrhenates

Neha Bura[1], Pablo Botella[1], Catalin Popescu[2], Frederico Alabarse[3], Ganapathy Vaitheeswaran[4], Alfonso Muñoz[5], Brendan J. Kennedy[6], Jose Luis Rodrigo Ramon[1], Josu Sanchez-Martin[1], Daniel Errandonea[1,*]

[1]Departamento de Física Aplicada - Instituto de Ciencia de Materiales, Matter at High Pressure (MALTA) Consolider Team, Universidad de Valencia, Edificio de Investigación, C/Dr Moliner 50, 46100 Burjassot, Valencia, Spain

[2]CELLS-ALBA Synchrotron Light Facility, Cerdanyola del Vallès, 08290 Barcelona, Spain

[3]Elettra Sincrotrone Trieste, Trieste, 34149, Italy

[4]School of Physics, University of Hyderabad, Prof. C. R. Rao Road, Gachibowli, Hyderabad 500046, Telangana, India

[5]Departamento de Física, MALTA Consolider Team, Universidad de La Laguna, San Cristóbal de La Laguna, Tenerife E-38200, Spain

[6]School of Chemistry, The University of Sydney, Sydney, New South Wales 2006, Australia

* daniel.errandonea@uv.es

**Abstract**

The effects of pressure on the crystal structure of scheelite-type perrhenates were studied using synchrotron powder X-ray diffraction and density-functional theory. At ambient conditions, the studied materials $AgReO_4$, $KReO_4$, and $RbReO_4$, exhibit a tetragonal scheelite-type crystal structure described by space group $I4_1/a$. Under compression, a transition from scheelite-to-M´-ferguosonite (space group $P2_1/c$) was observed at 1.6 and 7.4 GPa for $RbReO_4$ and $KReO_4$, respectively. The transition involves a relative volume decrease. On the other hand, $AgReO_4$ underwent a phase transition to the M-ferguosonite structure (space group $I2/a$) at 13.6 GPa. In this case there is no appreciable volume discontinuity. The room-temperature pressure-volume equation of state for the three studied perrhenates was estimated using a second-order Birch-Murnaghan equation of state. The results for the low-pressure phase are confirmed by density-functional theory calculations. The analysis of the bulk modulus shows that the compressibility of the compounds decreases following the sequence $RbReO_4 > KReO_4 > AgReO_4$, which is related to the compressibility of the $RbO_8$, $KO_8$, and $AgO_8$ bidisphenoid units. Density-functional theory also offers valuable insights into the elastic constants. Despite giving a good description for the low-pressure phase in the three compounds, density-functional theory cannot catch the structural phase transition observed in experiments. Reasons for it are discussed in the manuscript.

## 1. Introduction

Bimetal oxides form a large family of compounds, encompassing a wide range of materials. One of such group of compounds is formed by $AMO_4$ oxides which has received significant attention due to their multiple technological applications [1]. In this work, we focus on three specific members of the perrhenate family, $KReO_4$, $RbReO_4$, and $AgReO_4$. These compounds exhibit, at ambient conditions, a tetragonal scheelite-type structure with space group $I4_1/a$ which is schematically represented in Fig. 1. This structure has two types of polyhedra, $AO_8$ dodecahedra (with $A$ = K, Rb, Ag) and $ReO_4$ tetrahedra. In this structure, the $A$, Re, and O atoms occupy the Wyckoff positions 4a, 4b, and 16f.

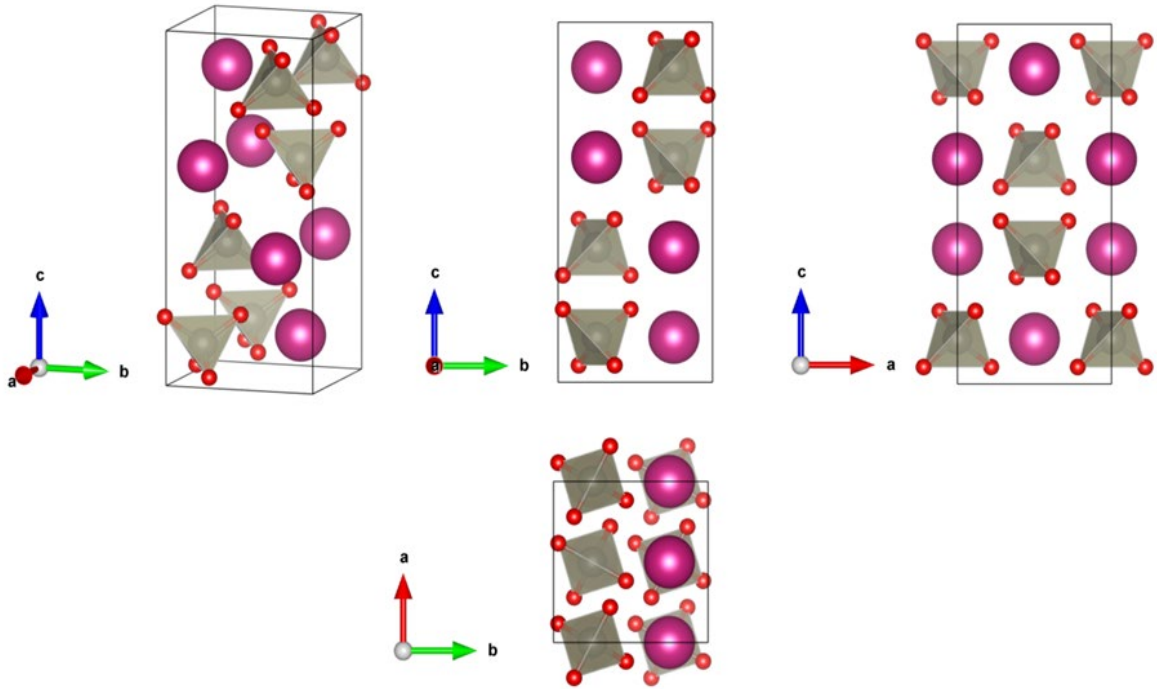

**Figure 1:** Different projections of the tetragonal scheelite-type crystal structure, described by space group $I4_1/a$. Spheres drawn with pink and silver colors represent $A$ (K, Rb, or Ag) atoms and Re atoms, respectively. Oxygen atoms are represented by red spheres. The $ReO_4$ tetrahedral units are shown.

Perrhenates have garnered significant interest in recent years due to their unique structural and electronic properties and potential applications across various fields, which include calorimeters for neutrino mass measurements [2]. Their fascinating electronic and optical behaviors make them suitable for use in optoelectronic devices, sensors, photoluminescence, and catalysts [3, 4]. Despite their potential, the study of $A$ReO$_4$ compounds under variable temperature and pressure conditions is relatively limited compared to related materials such as vanadates [5] and tungstates [1], highlighting the need for further exploration of the high-pressure properties of $A$ReO$_4$ perrhenates to fully understand and harness their functional potential.

The room-temperature pressure dependent Raman studies for KReO$_4$ of Jayaraman *et al.* showed a sequence of transitions at 7.5, 10, and 14 GPa [6]. Based on the changes in Raman spectra, the transitions were proposed to gradually reduce the symmetry of the crystal from tetragonal to orthorhombic to monoclinic to triclinic. Similar changes in the Raman spectra were detected at 1.6, 5.5, and 15 GPa for RbReO$_4$ [6]. Chay *et al.* studied the temperature dependent behaviour of RbReO$_4$ and observed a transition from the scheelite structure ($I4_1/a$) to another tetragonal structure described by space group $I4_1/amd$ near 650 K [7]. On the other hand, using energy-dispersive X-ray diffraction (XRD), as well as Raman spectroscopy, Otto *et al.* observed a phase transition in AgReO$_4$ to an unknown structure at 13 GPa [8, 9]. More recently, Mukherjee *et al.* studied the high-pressure (HP) properties of AgReO$_4$ using density-functional theory (DFT) simulations [10]. These authors suggested, based on the pressure dependence of the volume, that the previous XRD results [8] were hindered by non-hydrostatic effects. Critically, the crystal structure of the high-pressure phases of AgReO$_4$, KReO$_4$, and RbReO$_4$ have not yet been determined. Evidently it is timely to perform HP XRD studies of these compounds to determine the crystal structure of the HP phases using state-of-the-art methods.

To fully characterize the behaviour of these materials under extreme conditions, in this paper we concentrate on their crystal structures. We performed HP XRD experiments and focused on the determination of the crystal structure of different phases, as well as the pressure dependence of unit-cell parameters and the accurate determination of the room-temperature pressure-volume equation of state (EoS). We also performed DFT studies using different functionals to find the most appropriate functionals to describe the properties of the low-pressure phase in the different studied compounds.

## 2. Materials and Method

### 2.1 Sample synthesis and characterization

Polycrystalline samples of $KReO_4$, $RbReO_4$, and $AgReO_4$ were prepared as described by Chay *et al.* [6] A $HReO_4$ solution (100 mL, 53.7 mmol) was prepared by mixing rhenium metal (Aldrich, 99.9%,) 11.59 mmol, with 100 mL of fresh $H_2O_2$ (Sigma-Aldrich, 30% v/v) at room temperature. The solution was stirred overnight until the Re metal had completely dissolved. The target compound was then prepared by adding $A_2CO_3$, $A$ = K, Rb, Ag, (30 mL, 1.223 M) to the $HReO_4$ solution (100 mL, 0.0537 M) and allowing the product to precipitate. This was collected by filtration, washed with cold water and air dried.

Powder XRD studies were carried out to identify the phase of the samples along with their crystal structure. These measurements were performed with Cu K$\alpha_1$ radiation on a Rigaku Ultima IV diffractometer and Rietveld refinements were carried out using FULLPROF [11]. The three compounds were found to exhibit the tetragonal scheelite-type structure. The refined lattice parameters at ambient conditions are summarized in Table 1 and are in good agreement with the values previously reported by Chay *et al.* [7].

### 2.2 High-pressure studies

Angle dispersive XRD (AD-XRD) studies on $KReO_4$ and $RbReO_4$, to pressures of 10 GPa, were carried out at the Xpress beamline of the Elettra Synchrotron Radiatiom Facility (Elettra). AD-XRD studies on $AgReO_4$ were carried out up to 18.5 GPa at the MSPD beamline of the ALBA synchrotron [12]. The monochromatic wavelength for the experiments at Elettra was tuned at 0.4956 Å, and for experiments at ALBA it was 0.4246 Å. All the experiments were carried out using a membrane-type diamond-anvil cell (DAC) with a culet size of 500 μm. The samples were loaded in stainless steel gaskets, into which a 150 μm hole had been drilled, along with a small amount of Cu or Ag powder (used for pressure determination) and a 4**:**1 methanol-ethanol mixture, which acts as the pressure-transmitting medium (PTM). At each pressure, we collected two XRD patterns, one with Cu (or Ag) and sample used to determine the pressure, and one where we maximized the sample signal which was used for structural analysis. Pressure was determined with the equation of state of Cu (or Ag) reported by Dewaele *et al.* [13] with a 0.5% accuracy. The selected PTM remains quasi-hydrostatic up to 10 GPa [14] and is commonly used to study oxides in the pressure range covered by this study [15]. This medium is the same as that used in previous experiments [7, 8, 9], which allows a direct comparison of results. A PILATUS3 S 6M (Rayonix SX165 CCD) detector was used to collect the diffraction patterns in Elettra (ALBA). The detectors were calibrated using $LaB_6$ (ALBA) and $CeO_2$ (Elettra) as standards. To obtain a conventional 1-D diffraction pattern, the intensity was integrated as a function of $2\theta$ using Dioptas [16].

**Table 1:** Lattice parameters and volume of perrhenates obtained from XRD measurements at ambient conditions and determined from DFT calculations. For DFT calculations, we show in brackets the relative difference with the present experiments. Results from previous experiments are shown for comparison.

| Material | Space group $I4_1/a$ | Experiments | | DFT | | |
|---|---|---|---|---|---|---|
| | | This work | Chay *et al.*[7] | MetaSCAN | PBEsol | PBEsol + D3+BJ |
| $KReO_4$ | a (Å) | 5.6738(1) | 5.67625(8) | 5.6907 (0.30%) | 5.74597 (1.26%) | 5.74527 (1.24%) |
| | c (Å) | 12.6961(3) | 12.6994(4) | 12.5989 (-0.77%) | 12.5628 (-1.06%) | 12.78474 (0.69%) |
| | V (Å³) | 408.71(1) | 409.17(1) | 408.00 (-0.17%) | 414.77 (1.46%) | 422.00 (3.15%) |
| $RbReO_4$ | a (Å) | 5.8327(3) | 5.8329(1) | 5.8592 (0.45%) | 5.90404 (1.21%) | 5.91768 (1.44%) |
| | c (Å) | 13.2578(8) | 13.2543(3) | 13.0497 (-1.59%) | 13.04719 (-1.61%) | 13.24997 (-0.06%) |
| | V (Å³) | 451.04(4) | 450.94(2) | 448.00 (-0.68%) | 454.80 (0.83%) | 464.00 (2.79%) |
| $AgReO_4$ | a (Å) | 5.3664(6) | 5.37674(7) | 5.3375 (-0.54%) | 5.33738 (-0.54%) | 5.369 (0.05%) |
| | c (Å) | 11.8450(31) | 11.8006(2) | 11.5208 (-2.81%) | 11.32421 (-4.60%) | 11.795 (-0.42%) |
| | V (Å³) | 341.11(9) | 341.15(1) | 328.00 (-4.0%) | 322.60 (-5.74%) | 340.01 (-0.33%) |

## 2.3 Density-functional theory calculations

First-principles calculations were performed utilizing the well-established plane-wave pseudopotential method within the context of DFT, as implemented in the Vienna Ab initio Simulation Package (VASP). Calculations for the low-pressure phase of AgReO$_4$ were already published [10]. It that work we have shown that the generalized-gradient approximation (GGA), using PBEsol for the exchange-correlation functional including van der Waals correction using the D3 method proposed by Grimme, incorporating the Becke–Johnson (BJ) damping variant (D3+BJ) was the most accurate method to describe the scheelite phase of AgReO$_4$ [10]. Here, for KReO$_4$ and RbReO$_4$, we compare calculations using PBEsol and PBEsol including D3+BJ. We also compared these functionals for the three compounds with the newly developed strongly constrained and appropriately normed family of meta-GGA density functionals (MetaSCAN) [17]. MetaSCAN was also used in this work to study AgReO$_4$ for the sake of completeness. Calculations with the HSE06 [18] hybrid functionals gave very similar results to MetaSCAN.

In all calculations the projector augmented wave (PAW) pseudopotentials from the VASP database were utilized. The valence electron configurations were defined as follows: for K, [Ar] $4s^1$; Rb, [Kr] $5s^1$; Ag, [Kr] $4d^{10}$ $5s^1$; Re, [Xe] $4f^{14}$ $5d^5$ $6s^2$; and O, [He] $2s^2$ $2p^4$. To enhance computational precision, a dense Monkhorst–Pack sampling of a 9 × 9 × 9 k-point mesh was employed for Brillouin zone integration. A plane-wave basis set with an energy cut-off of 600 eV was selected to ensure accurate and well-converged structural results. The criteria for self-consistency in energy convergence were established at $1 \times 10^{-8}$ eV per atom, while the maximum interatomic force was limited to 0.002 eV/Å. Stability criteria were established to optimize structural parameters across different volumes and pressures, ensuring that the deviation of the stress tensor from a diagonal hydrostatic form remained below 0.1 GPa. From these calculations, a dataset comprising the lattice parameters of the ground state, energies (E) and volumes (V) at varying pressures (P) derived from the stress tensor was generated. This dataset was fitted using the third-order Birch–Murnaghan EoS to extract the bulk modulus and its pressure derivative. As shown in Table 1, we found that the PBEsol+D3+BJ functional best describes the ground states of AgReO$_4$, but MetaSCAN works better for KReO$_4$ and RbReO$_4$. We also computed the mechanical properties of the three compounds. Evaluation of the mechanical properties involved the calculation of elastic constants.

These constants were obtained from the stress tensor, which was determined by applying strain to the relaxed structure through alterations in its lattice vectors, encompassing both magnitude and angle, using the stress-strain approach implemented in VASP. The elastic moduli were derived from the elastic constants.

## 3. Results and Discussion

### 3.1 Phase Transitions

#### 3.1.1 Potassium Perrhenate (KReO$_4$)

High-pressure powder XRD studies on KReO$_4$ were carried out to up to 10 GPa. A selection of XRD patterns at different pressures is shown in Fig. 2 including Rietveld refinements. As previously mentioned, the material has a tetragonal scheelite-type structure ($I4_1/a$) at ambient conditions. This structure persists to around 6.4 GPa. The XRD pattern measured at 7.1 GPa contains a few additional peaks, most obvious around $2\theta \sim 6°$, and scrutiny of this suggests the coexistence of two phases. At higher pressures no peaks diagnostic of the tetragonal scheelite phase could be observed, rather the diffraction patterns could be fit to a monoclinic phase. Analysis suggested this high-pressure phase has a primitive monoclinic cell and is isomorphic to the so-called M'-fergusonite structure described in space group $P2_1/c$. This structure has previously been identified as a HP phase in HoNbO$_4$ and in double molybdates [19, 20]. A two phase ($I4_1/a$ and $P2_1/c$) model was generated and this provided a good fit to the profile measured at 7.1 GPa. The diffraction patterns measured at and above 7.4 GPa only displayed peaks corresponding to the HP monoclinic M'-fergusonite phase.

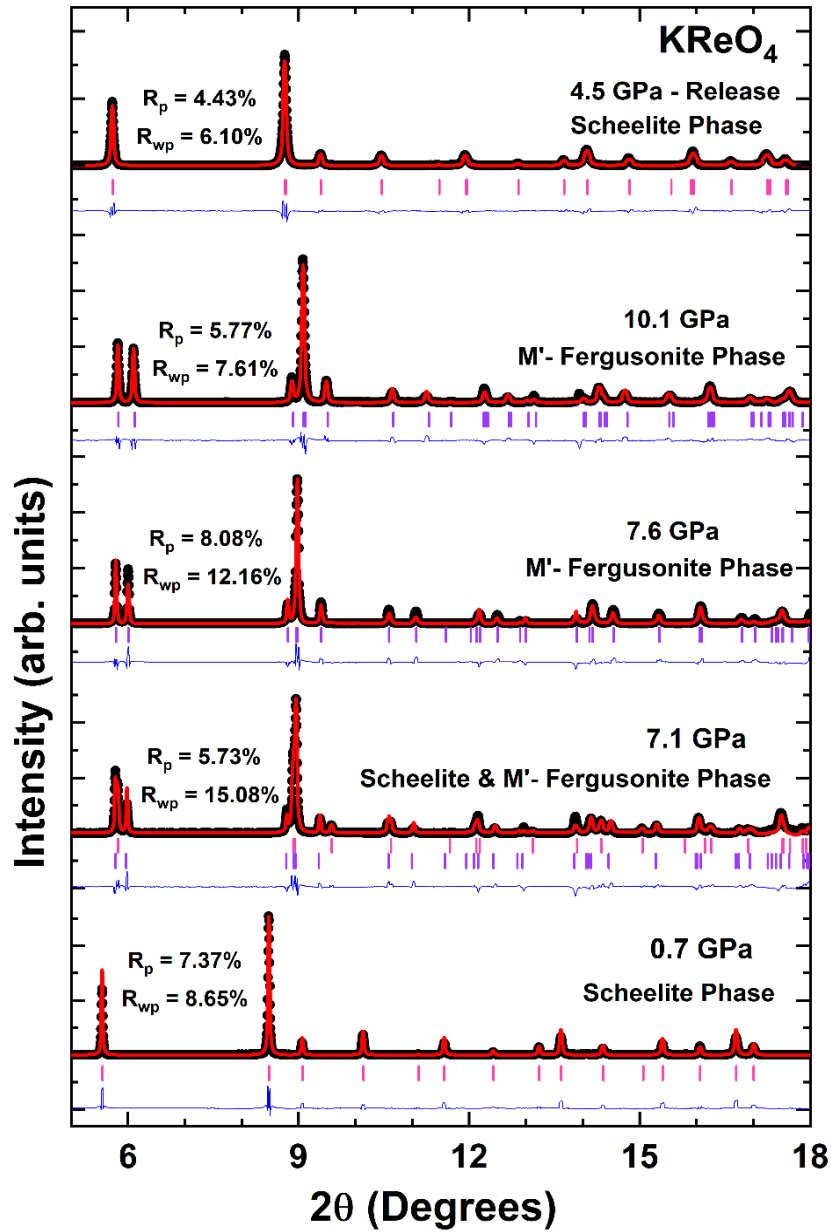

**Figure 2:** Rietveld refinements of XRD pattern of KReO$_4$ at selected pressures, $\lambda$ = 0.4956 Å. Pink (Purple) ticks identify the positions of Bragg reflections of the low-pressure tetragonal (high-pressure monoclinic) phase. The black circle represents the experimental data, the calculated profile is given by the red lines, and the blue lines represent the difference between the calculated and measured profiles. R-values of the refinements are included in the figure.

The HP structure is represented in Fig. 3. The lattice parameters at 7.4 GPa are summarized in Table 2. The HP monoclinic phase is found to be stable up to 10.1 GPa. Given the coexistence of the low pressure tetragonal and high-pressure monoclinic phases at 7.1 GPa and the volume discontinuity associated with the transition (see section 3.2) the phase transition is evidently first-order. The phase transition is found to be reversible, as the peaks observed at low pressure during, the decompression cycle, were found to belong to the scheelite phase (see Figure 2). The M'-fergusonite structure is a distorted and compressed form of the scheelite structure. This structural change is due to small deformations of the cation matrix and significant displacements of the anions. The most noticeable change after the transition are the discontinuous decrease of the unit-cell volume and the distortion of $KO_8$ and $ReO_4$ polyhedra and the way that polyhedra are interconnected. At the phase transition the discontinuity of the unit-cell volume is approximately 1.2 %. On the other hand, the distortion index defined by Baur [21] changes from 0.0174 in scheelite at 7.14 GPa to 0.0552 in M'-fergusonite at 7.44 GPa. IThe $ReO_4$ tetrahedra is regular in the low-pressure phase with four identical bond distances, and distorted in the M'-fergusonite with a distortion index [21] of 0.0078.

**Table 2:** Refined lattice parameters and atomic positions of the M'-fergusonite phase of $KReO_4$ at 7.4 GPa.

| Parameters | Values | Atoms | x | y | z | Wyckoff Position |
|---|---|---|---|---|---|---|
| Space Group | $P2_1/c$ | K | 0.7508(3) | 0.1265(3) | 0.0013(3) | 4e |
| a (Å) | 5.1461(5) | Re | 0.7498(3) | 0.6255(3) | 0.0017(3) | 4e |
| b (Å) | 12.1067(9) | O1 | 0.9714(7) | 0.7037(4) | 0.1224(6) | 4e |
| c (Å) | 5.3689(4) | O2 | 0.5283(6) | 0.7040(7) | 0.8761(7) | 4e |
| β (°) | 90.09(1) | O3 | 0.1271(5) | 0.4535(6) | 0.2204(6) | 4e |
| Volume (Å³) | 334.49(5) | O4 | 0.3723(6) | 0.4543(8) | 0.7781(6) | 4e |

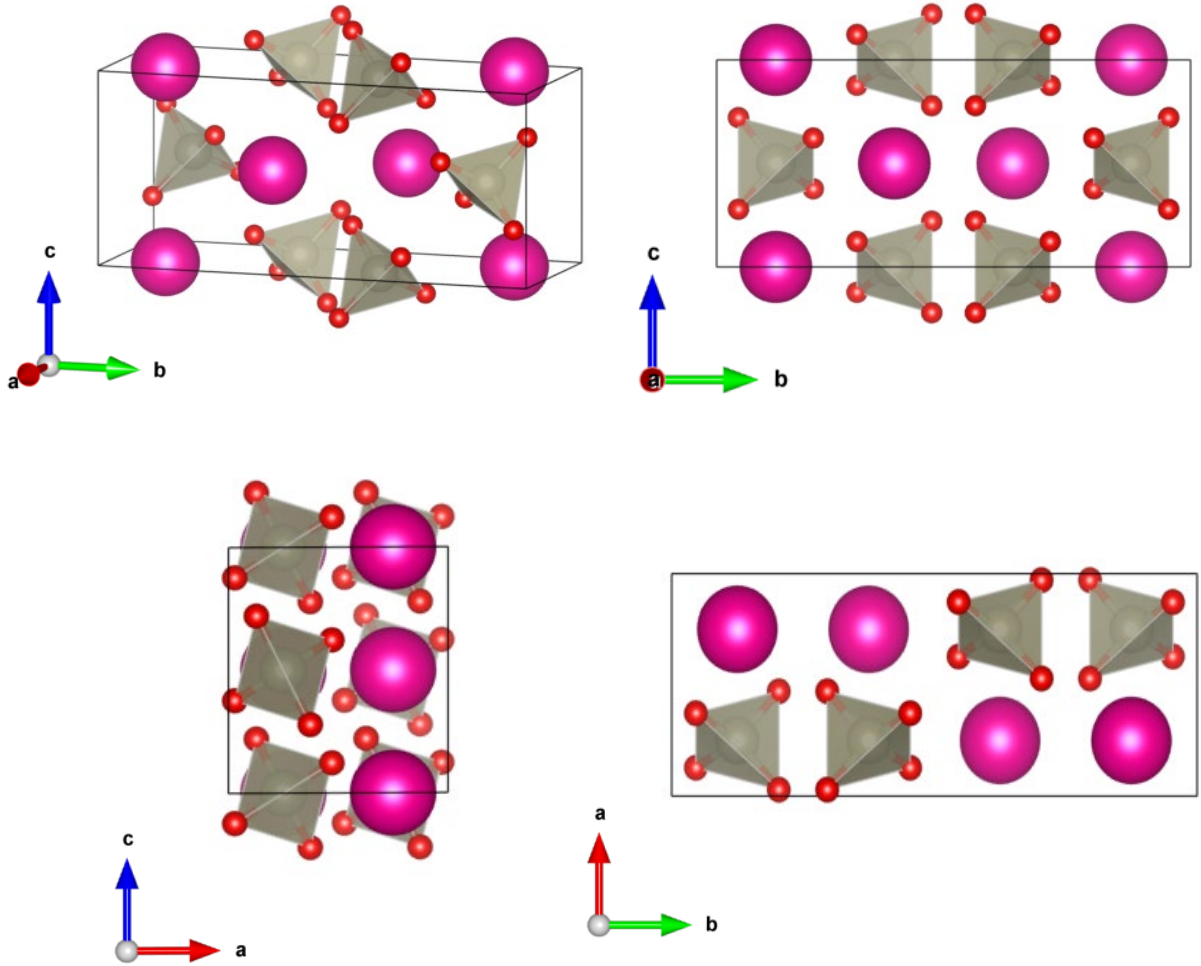

**Figure 3:** (a) Representation of the high-pressure monoclinic crystal structure of perrhenates in space group $P2_1/c$. The pink spheres represent $A$ (K or Rb) atoms. The grey spheres represent the Re atoms and the oxygen atoms are represented by red spheres. The ReO$_4$ polyhedra are shown.

The phase transition observed in KReO$_4$ is in keeping with the trends described by Jayaraman *et al.*, who reported phase transitions at 7.5 and 10 GPa [6]. The monoclinic HP phase we are proposing involves an increase of the number of Raman modes compared to low-pressure scheelite structure. This is consistent with the changes observed in the Raman spectrum by Jayaraman *et al.* [6] at the phase transition. DFT calculations were performed taking as starting models the HP monoclinic phase obtained from the current experiments. Invariably, when optimizing the HP monoclinic structure, it always reverted to the tetragonal scheelite structure, which is facilitated by the fact that both structures are related by group-subgroup relations ($P2_1/c \subset C2/c \subset I4_1/a$). Surprisingly irrespective of the functional employed in our calculations, PBEsol, PBEsol+D3+BJ, HSE06, MetaScan, PBE, and AM05, the DFT calculations failed to reproduce the observed transition. This discrepancy could be related to the influence of non-hydrostatic effects in experiments, which could favor the formation of metastable phases at pressures where they are not thermodynamically stable [22]. However, we are confident that this is not the case in our study for two reasons. The transition found in our XRD measurements happens at a similar pressure to that reported in previous experiments [6]. When methanol-ethanol is used as a pressure medium, as in our and previous studies, non-hydrostaticity only becomes noticeable beyond 10 GPa [14], i.e. at pressures higher than the transition pressure. We consider, therefore, that the discrepancy might be related to the fact that DFT is not capturing specific features of the phase transition, such as a possible pressure-induced delocalization of $f$-electrons of Re [23], which could strongly affect the HP behaviour of materials. It is remarkable that a similar discrepancy was also found in this study for $RbReO_4$ and $AgReO_4$, see below. It is hypothesized that this might be related to a poor description of the Re $f$-electrons under HP. Under compression, the interatomic distances decrease, further increasing the strength of $f$-electrons correlation, a phenomenon that often DFT does not accurately capture [24]. Another example of the problems of DFT for describing rhenium compounds under HP is $ReO_3$, for which DFT underestimates the pressure induced changes in the unit-cell volume below 10 GPa [25].

### 3.1.2 Rubidium Perrhenate ($RbReO_4$)

Powder XRD studies on $RbReO_4$ were also carried out up to a pressure of 10 GPa. A selection of XRD patterns, including Rietveld refinements, is shown in Fig. 4. A similar tetragonal to monoclinic transition to that described for $KReO_4$ was observed around 1.0 GPa. At this pressure, extra peaks started appearing in the diffraction pattern, most obvious near $2\theta \sim 5.5$ and $8.0°$ Although Jayaraman *et al.,* [6] based on Raman spectroscopy, suggested that the first HP structure in $RbReO_4$ has the same orthorhombic structure observed in $CsReO_4$ and $TlTcO_4$ [26, 27] the current diffraction data shows that it actually has a monoclinic structure and is isostructural to the HP monoclinic $P2_1/c$ phase that forms in $KReO_4$ above 7 GPa.

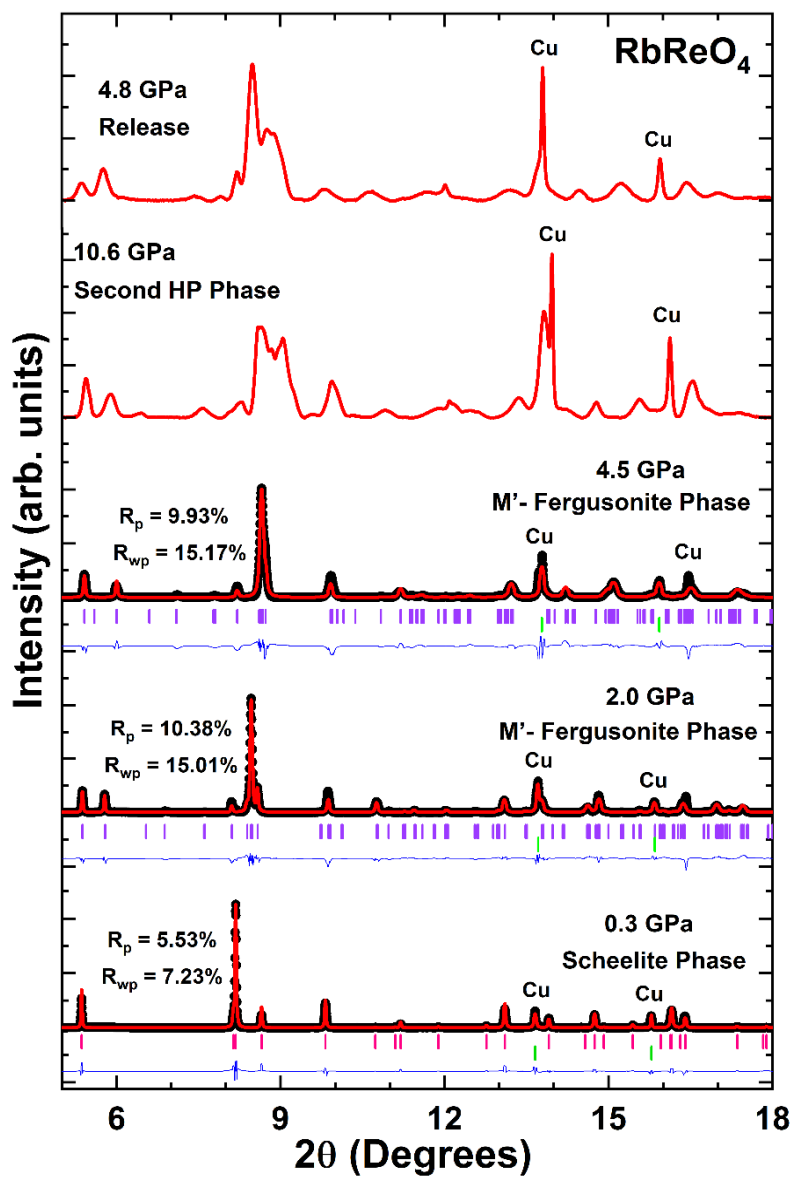

**Figure 4:** Rietveld refinements of XRD pattern of RbReO$_4$ at 0.3, 2.0, and 4.5 GPa, $\lambda$ = 0.4956 Å. The pink and purple ticks identify the position of the Bragg peaks of the low-pressure tetragonal and HP monoclinic phase, respectively. The green peaks show the position of the Cu peaks. The black circles represent the measured data, the calculated profiles are given by the red lines, and the blue lines represent the difference between them. Cu peaks are identified. The figure also includes the XRD patterns measured at 10.6 GPa showing evidence of a second phase transition to an unidentified phase and at 4.8 GPa under pressure release. The changes in the intensity of Cu peaks at these two pressures indicate a shift in the preferred orientation of crystal planes of Cu. R-values of the refinements are included in the figure.

Table 3 gives the lattice parameters corresponding of RbReO$_4$ at 2.0 GPa. This HP primitive monoclinic phase of RbReO$_4$ remains stable to pressures of up to 4.5 GPa. In this compound the discontinuity of the unit-cell volume at the transition is approximately 2.1%. On the other hand, the distortion index [21] of the RbO$_8$ polyhedron changes from 0.0144 to 0.0418 and the ReO$_4$ tetrahedron changes from regular to being irregular with a distortion index [21] of 0.0068. Beyond 4.5 GPa a second transition occurs, however, as is obvious from Fig. 4 the diffraction peaks from this second HP phase are severely broadened. That this is not a result of a loss of hydrostatic conditions is evident from the well resolved peaks from the Cu pressure standard. The peaks corresponding to the second HP phase have very low intensity and as is evident from Fig. 4 this is not the primitive monoclinic M'-fergusonite phase. Due to the relatively poor quality of the XRD patterns measured at pressure above 4.5 GPa, that further decreased as the pressure increased, it was not possible to solve the structure of this phase.

**Table 3:** Refined lattice parameters and atomic positions of the M'-fergusonite phase of RbReO$_4$ at 2.0 GPa.

| Parameters | Values | Atoms | x | y | z | Wyckoff Position |
|---|---|---|---|---|---|---|
| Space Group | $P2_1/c$ | Rb | 0.7503(4) | 0.1246(4) | 0.0007(4) | 4e |
| a (Å) | 5.2854(5) | Re | 0.7505(4) | 0.6252(4) | 0.0011(4) | 4e |
| b (Å) | 13.2462(9) | O1 | 0.9712(8) | 0.7039(8) | 0.1227(6) | 4e |
| c (Å) | 5.7556(4) | O2 | 0.5288(7) | 0.7039(8) | 0.8773(9) | 4e |
| β (°) | 90.15(1) | O3 | 0.1273(6) | 0.4539(7) | 0.2212(5) | 4e |
| Volume (Å³) | 402.96(5) | O4 | 0.3727(7) | 0.4539(7) | 0.7780(7) | 4e |

Our findings are consistent with the results reported by Jayaraman *et al.* [6], who reported transitions in RbReO$_4$ at 1.5 and 5.5 GPa, based on Raman studies. The changes in the Raman spectra at the first transition [6], included an increase in the number of Raman modes that, as mentioned above, is consistent with the tetragonal-monoclinic phase transition proposed here. Unlike in KReO$_4$, in RbReO$_4$ the observed phase transitions are not reversible, at least when pressure is reduced to 4.8 GPa. Fig. 4 shows that the XRD pattern measured at 4.8 GPa in the decompression cycle resembles that measured at 10.6 GPa during compression. Unfortunately, we could not collect XRD at lower pressures during decompression because of friction between the piston and cylinder of the DAC; the lowest pressure obtainable removing all the force applied to the DAC was 4.8 GPa. Since this pressure is higher than the transition pressure of the second phase transition, we cannot extract from present XRD experiments any conclusion of the reversibility of the phase transitions of RbReO$_4$. As found for KReO$_4$, DFT cannot capture the transition from scheelite to M'-fergusonite, in RbReO$_4$. That the transition happens around 1 GPa effectively rules out the possibility that the observed results are impacted by non-hydrostatic effects.

### 3.1.3 Silver Perrhenate (AgReO$_4$)

A selection of XRD patterns for AgReO$_4$ measured at different pressures is given in Fig. 5. The XRD studies revealed a systematic shift of peaks to higher 2θ upon application of pressure, which can be attributed to compression of the lattice. At 13.63 GPa the most intense peak around 2θ ~ 7.7 ° begins to split into two distinct peaks. Similar splitting occurs for other peaks. This splitting of peaks is enhanced at higher pressure. Splitting of the scheelite 101, 112/203, and 200 peaks is clearly seen in the XRD patterns measured 13.9 and 18.4 GPa as illustrated in Fig. 5. This suggests a monoclinic distortion in the (001) plane of scheelite. It should be stressed that no new peaks emerged in the diffraction patterns until the highest studied pressure.

A key observation here is the lack of additional reflections that shows the cell remains I-centered, and a satisfactory fit to the data measured at 13.6 GPa was obtained to the monoclinic fergusonite structure. The fergusonite is best described in the non-standard *I*2/*b* setting of space group *C*2/*c* to facilitate comparison with the tetragonal *I*4$_1$/*a* scheelite structure, but in the present case *I*2/a is preferred as it allows comparison with the primitive monoclinic (*P*2$_1$/*c*) M'- fergusonite structure seen for KReO$_4$ and RbReO$_4$ at high pressure. The HP phase remains stable up to the highest pressure covered by the present study, 18.5 GPa. These results agree with the earlier study by Otto *et al.* [8, 9] who reported a phase transition to occur at 13(1) GPa. The scheelite to M-fergusonite transition is a typical pressure-driven transition of scheelite-structured oxides [28, 29]. The two crystal structures are related via a group-subgroup relationship *I*2/a ⊂ *I*4$_1$/*a*. M-fergusonite can be obtained from scheelite via a shear deformation of the *xy* plane of the tetragonal structure and a slight displacement of the atoms, with no dramatic reconstruction of the lattice [30]. This transition is driven by a Γ-point soft optical B$_g$ phonon [31, 32]. As shown below, the phase transition does not involve any discontinuity in the pressure dependence of the volume.

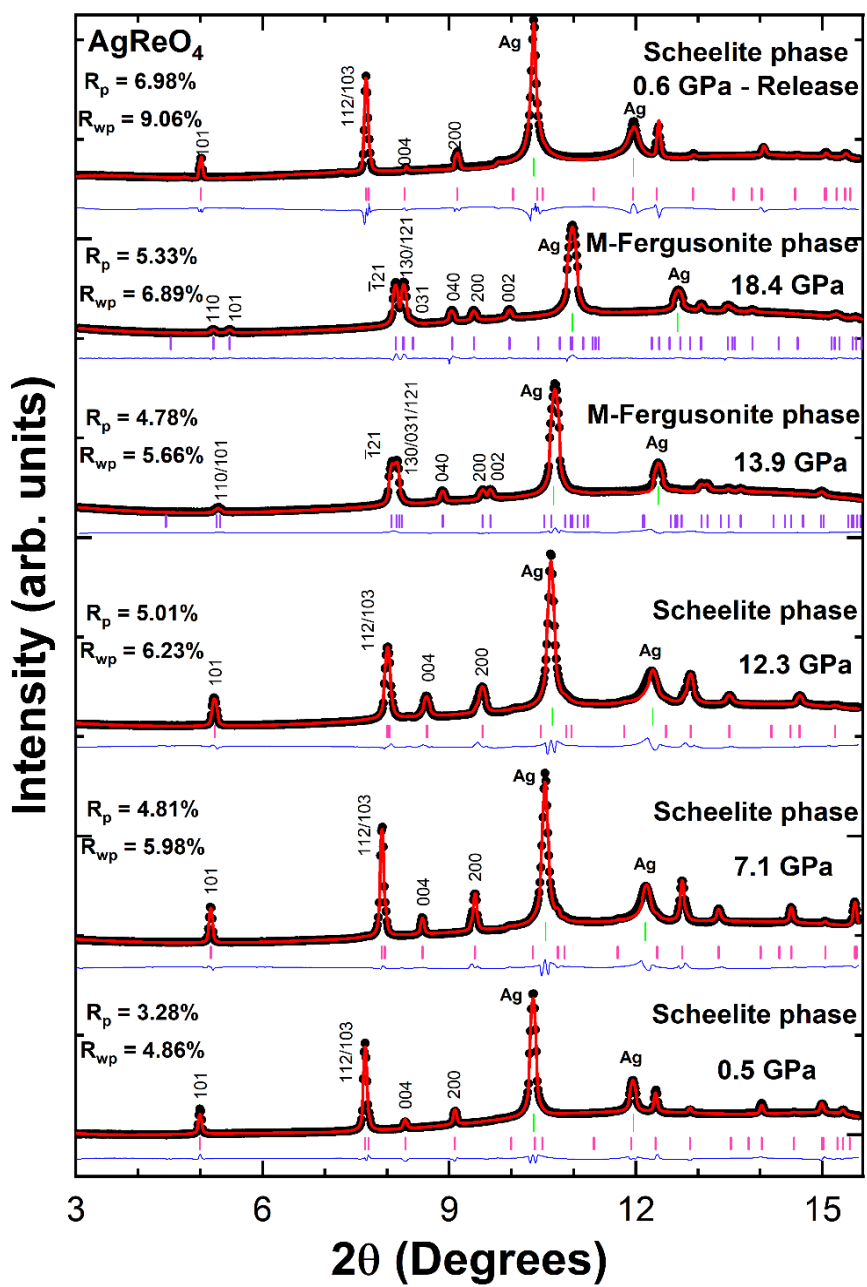

**Figure 5:** Rietveld refinements of XRD patterns of AgReO$_4$ at selected pressures, $\lambda$ = 0.4246 Å. The black circles represent the measured data, the calculated profiles are given by the red lines, and the blue lines represent the difference between them. The pink and purple tick markers identify the position of peaks of the low-pressure tetragonal scheelite and high-pressure M-fergusonite phases respectively. The green ticks indicate the position of Ag peaks. The Miller index relevant for the discussion are labeled, and Ag peaks are identified. The splitting of the 101, 112/103, and 200 peaks of scheelite is an indication of the phase transition. R-values of the refinements are included in the figure.

**Table 4:** Lattice parameters of the M-ferqusonite of AgReO$_4$ at 13.9 GPa.

| Parameters | Values | Atoms | x | y | z | Wyckoff Position |
|---|---|---|---|---|---|---|
| Space Group | *I2/a* | Ag | 0.25 | 0.1282(6) | 0 | 4e |
| a (Å) | 5.110(5) | Re | 0.25 | 0.6131(6) | 0 | 4e |
| b (Å) | 10.95(1) | O1 | -0.6178(23) | 0.0419(34) | -0.7694(29) | 8f |
| c (Å) | 5.762(5) | O2 | -0.5194(32) | 0.2919(27) | -0.6322(33) | 8f |
| β (°) | 90.81(9) | | | | | |
| Volume (Å³) | 282.1(4) | | | | | |

Table 4 reports the structural information of the crystal structure of the HP phase of AgReO$_4$. The structure of the M-ferqusonite phase is shown in Fig. 6. The HP M-ferqusonite phase proposed here has 18 Raman-active modes, compared to 13 Raman active modes in the tetragonal scheelite. Otto *et al.* [9] reported that the number of Raman modes increases at the tetragonal to monoclinic transition. The M-ferqusonite structure is also consistent with the splitting of the high-frequency stretching modes reported previously, which is a typical fingerprint of the scheelite to M-ferqusonite transition [33]. As observed for the other perrhenates studied here, the DFT calculations did not capture the phase transition. Using the M-ferqusonite structure as a starting model for the calculations, the structure invariably relaxed to the tetragonal scheelite upon optimisation. As discussed above, understanding this issue is beyond the scope of the present work. DFT did however accurately describe the pressure dependence of unit-cell parameters of the scheelite phase.

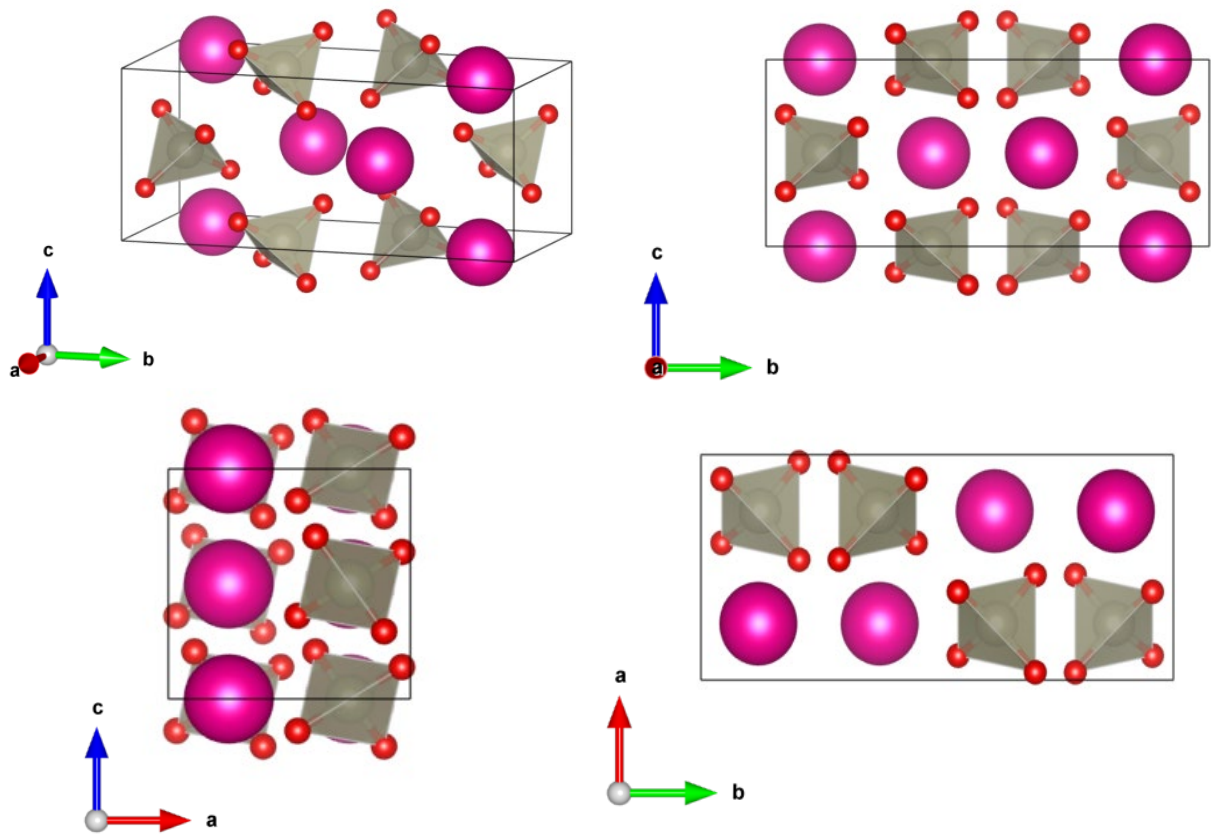

**Figure 6:** (a) Representation of the high-pressure monoclinic crystal structure of perrhenates in space group $I2/a$. The pink spheres represent Ag atoms. The grey spheres represent the Re atoms, and the oxygen atoms are represented by red spheres. The ReO$_4$ polyhedra are shown.

### 3.2 Pressure dependence of lattice parameters and unit-cell volume

Structural refinements against the XRD patterns provided accurate lattice parameters and unit-cell volume, which were used to calculate the bulk modulus ($K_0$) using a second-order Birch-Murnaghan EoS [34]. Fig. 7 illustrates the pressure dependence of the lattice parameters for the three studied oxides. As evident from this figure the pressure induced compression is anisotropic, reflecting the layered nature of the scheelite structure, with the compression along the $c$-axis being greater than in the $ab$-plane.

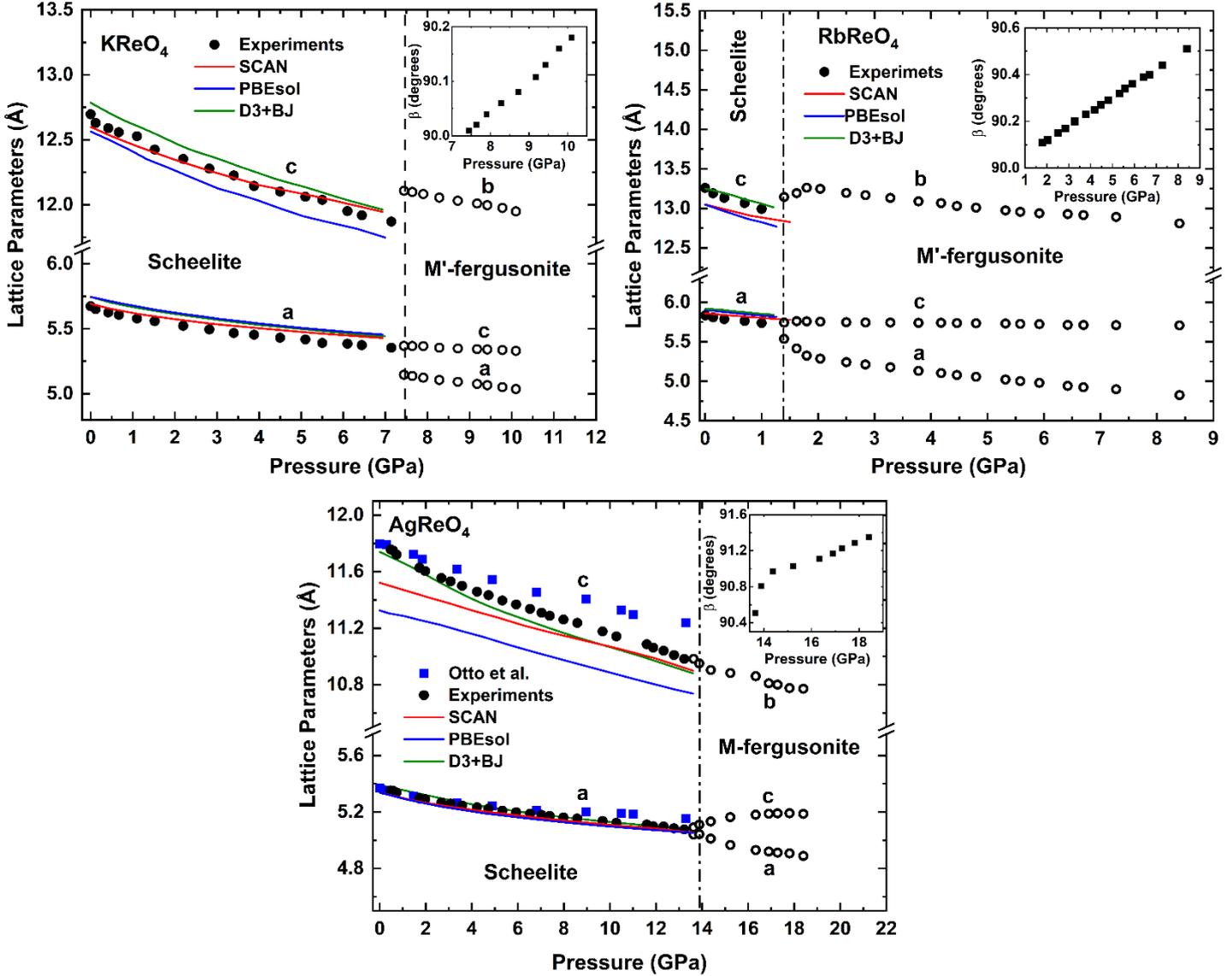

**Figure 7:** Pressure dependence of the lattice parameters for $A$ReO$_4$ ($A$ = K, Rb, Ag) oxides. The error bars are smaller than the size of the symbols. The inset of the graphs shows the change in the β-angle with pressure for the HP monoclinic phases. The results from the DFT calculations are shown as continuous lines. For AgReO$_4$, results from previous XRD experiments by Otto *et al.* [8] are included as blue symbols. The D3+BJ results for AgReO$_4$ are taken from reference [10]. The vertical lines show the transition pressures. Due to space limitations in this, and the subsequent figures, we use "SCAN" to describe the MetaSCAN functional.

Figure 7 also includes the compression of the three oxides calculated using DFT. For AgReO$_4$ and RbReO$_4$ PBEsol+D3+BJ gives the best agreement whereas for KReO$_4$, the MetaSCAN calculations provided the best agreement. In this figure the results for AgReO$_4$ are compared with previous studies [10]. Our results for AgReO$_4$ compare better with previous DFT calculations [10] than with the previous XRD experiments [8]. The previous XRD experiments underestimate the compressibility of both axes possibly due to the influence of non-hydrostatic stresses caused by the sample bridging between diamonds [10].

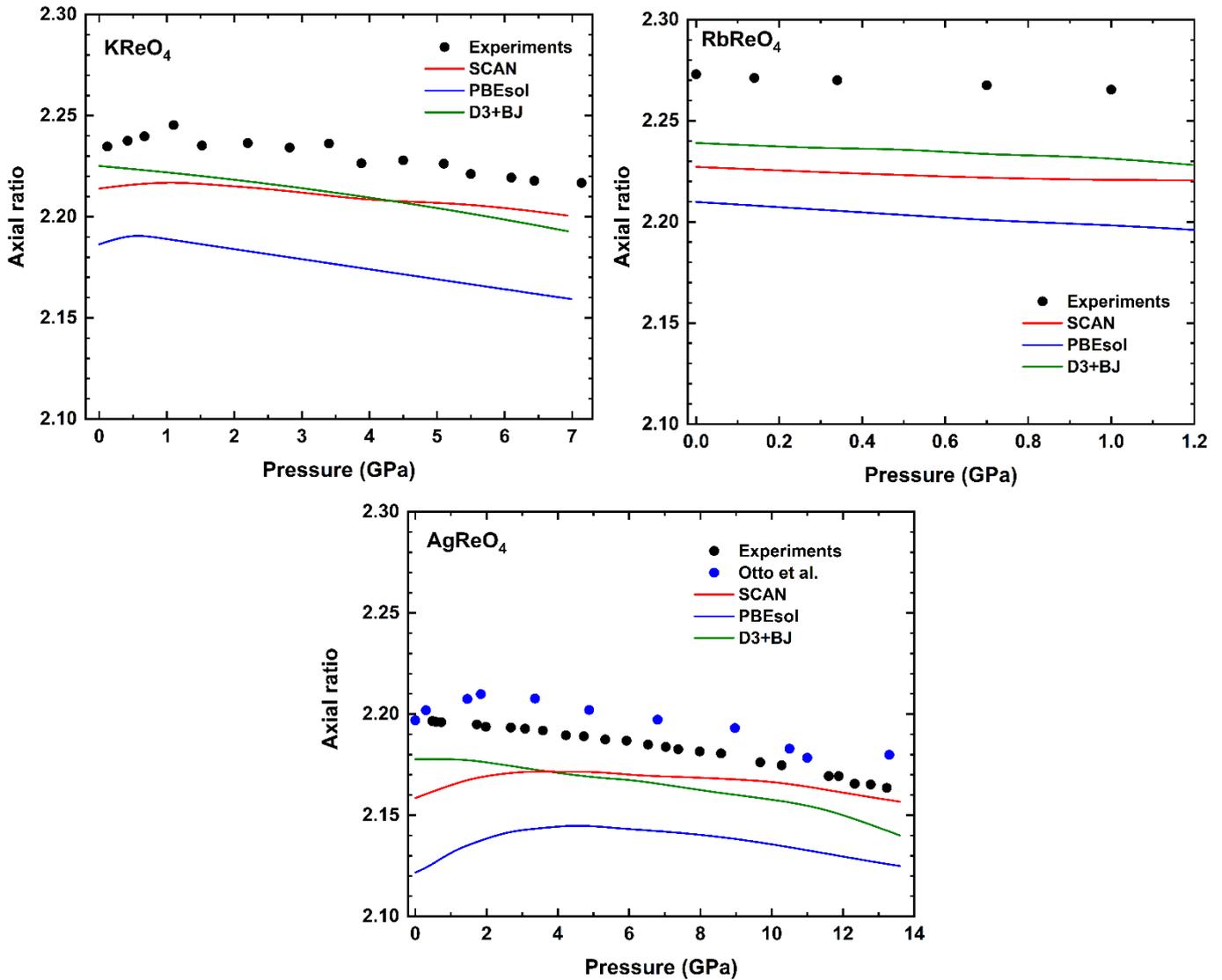

**Figure 8:** Pressure dependence of the axial ratio *c/a* ratio for perrhenates. The error bars are smaller than the size of the symbols. The results for the DFT calculations are also shown as solid lines. For AgReO$_4$ the black symbols are from this study and blue symbols from the experiments performed by Otto *et al.* [8].

In the three compounds, the scheelite phase is more compressible along the $c$-axis than along the $a$-axis. This anisotropic nature of compressibility can be seen clearly in the axial ratio vs pressure plot for the three perrhenates as shown in Fig. 8. At the phase transition, there is no discontinuity in the lattice parameters for AgReO$_4$, however, there is a clear discontinuity in KReO$_4$ and RbReO$_4$. In the HP monoclinic phases, the greatest compression is in the $b$-axis which corresponds to the $c$-axis of the tetragonal scheelite structure. We also observed that the monoclinic β angle increases with pressure in the three compounds and that the anisotropy of the basal $ac$ plane becomes greater as pressure increases, showing that the monoclinic distortion is enhanced by compression.

Fig. 9 presents the volume versus pressure curves for the three studied perrhenates and the fits to the second-order Birch-Murnaghan EoS. Both KReO$_4$ and RbReO$_4$ exhibit a first-order phase transition, characterized by a ~ 2.6% decrease in volume as they tranform from the tetragonal ($I4_1/a$) to the monoclinic ($P2_1/c$) phase. In contrast there is no obvious volume discontinuity at the $I4_1/a$ to $I2/a$ phase transition in AgReO$_4$. The scheelite to fergusonite transition observed for AgReO$_4$ is allowed to be continuous, reflecting the group-subgroup relationship between the two structures, although there is evidence that it is often not second order [35, 36]. However, a direct scheelite to M'-fergusonite transition cannot be continuous. The difference in behaviour can, in part, be attributed to the differences in the ionic radii of the $A$-type cations. According to the phenomenological model of $ABX_4$ structures proposed by Bastide [37] the fergusonite structure, that is derived from scheelite by a ferroelastic displacement of the cations coupled with rotation of the unconnected $BO_4$ tetrahedra, is favoured by smaller $A$-site cations such as Ag (1.28 Å). We propose that increasing the size of the $A$-site cation, as occurs for K (1.51 Å) and Rb (1.61 Å) restricts rotation of the tetrahedral units and the steric stresses induced by pressure are accommodated by deformation of the outer shell of the cation rather than simple rotation of the tetrahedra resulting in a reconstructive phase transition [38]. A similar trend to that seen here for the perrhenates has been observed in tungstates where, under pressure, CaWO$_4$ and SrWO$_4$ undergo the same $I4_1/a$ to $I2/a$ transition observed for AgReO$_4$ whereas BaWO$_4$ and PbWO$_4$ show a $I4_1/a$ to $P2_1/n$ transition, which was attributed to larger ionic radii of Ba and Pb than Sr and Ca [29, 39].

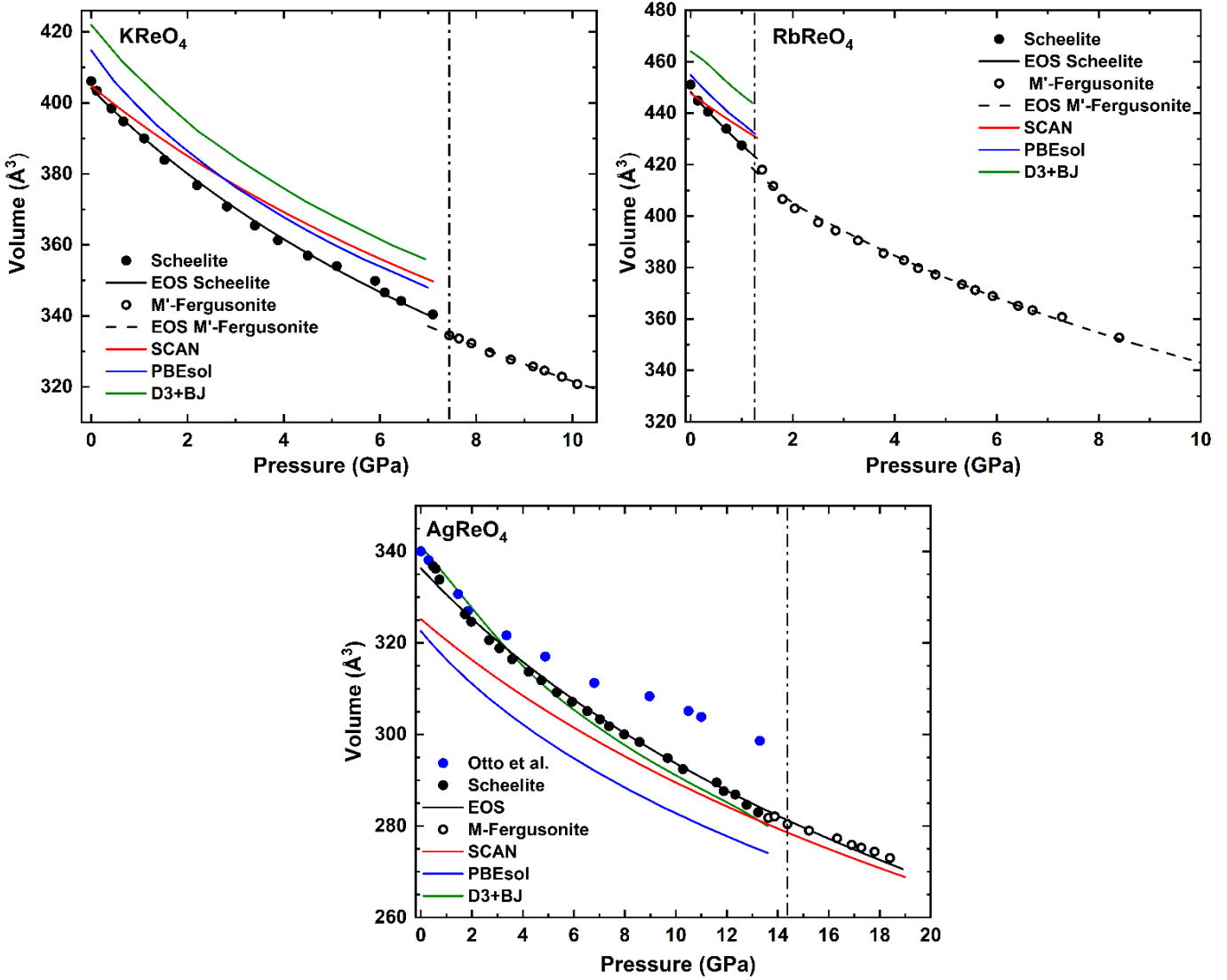

**Figure 9**: Pressure vs Volume data for perrhenates along with corresponding second-order Birch-Murnaghan fits, illustrating the volume discontinuity indicative of a 1$^{st}$ order phase transition in KReO$_4$ and RbReO$_4$. The error bars are smaller than the size of the symbols. The absence of such a discontinuity at the phase transition in AgReO$_4$ is indicative of a continuous transition. The figure for AgReO$_4$ includes results from an earlier XRD experiment (Otto *et al.* [8]). In all cases the closed symbols are for the tetragonal structures and the open symbols for the high-pressure monoclinic structures. The black lines represent the EoS fits described in the text, and colored lines represent the DFT calculations for the low-pressure phase. The D3+BJ calculations for AgReO$_4$ were previously published [10]. The vertical lines show the transition pressures.

The arguments presented in the previous paragraph combined with high-temperature studies allows us to obtain a systematic overview of the behavior of $A$ReO$_4$ perrhenates. It should be noted here that increasing temperature leads to an increase of the volume and increasing pressure leads to a decrease of the volume. So, both thermodynamic variables ae expected to have the opposite behavior in the structural sequence. It is known that CsReO$_4$, which has an orthorhombic structure related to scheelite (space group *Pnma*) and scheelite-type RbReO$_4$ (space group $I4_1/a$) transform at high-temperature to a crystal structure described by space group $I4_1/amd$ [7] which is derived from scheelite by removal of the tetrahedral rotation. On the other hand, a transition from a structure described by space $I4_1/amd$ to scheelite has been reported in many compounds under compression [40, 41]. Therefore, based on the comparable behavior of oxides with similar structures under both variable temperature and pressure conditions and the arguments proposed by Bastide, [37] we would expect that under compression CsReO$_4$ would transform to a M'-fergusonite structure as observed for RbReO$_4$. This hypothesis is consistent with the fact that the Raman spectra of the HP phases of CeReO$_4$ and RbReO$_4$ are similar [6]. On the other hand, given Na has an ionic radius more similar to that of Ag than that of K and Rb, we would expect that NaReO$_4$ would transform to the M-fergusonite structure under high-pressure. Both hypotheses should be confirmed by future HP XRD studies.

The results obtained for the pressure dependence of the volume of AgReO$_4$ are in good agreement with the present and previous calculations [10]. As noted elsewhere the present experimental results diverge from the previously reported values [9] around 2 GPa. It is postulated that the previous XRD studies were hindered by non-hydrostatic stresses. The values of bulk modulus obtained for each compound in this work are summarized in Table 5. This table also includes the values estimated for the high-pressure phases. In all cases, a second-order EoS was used (pressure derivative of the bulk modulus $K'_0 = 4$) to facilitate a direct comparison between results; in a third-order EoS $K_0$ and $K'_0$ are correlated, so the bulk moduli cannot be directly compared. The second-order EoS fits well the experimental results.

**Table 5:** Comparison of Bulk Moduli ($K_0$) between various theoretical approaches and experiments. $K_0$ is given in GPa. The volume at zero pressure ($V_0$) is included to provide the complete information of the different EoS. $V_0$ is given in Å$^3$.

|  | Low Pressure Phase | | | | High Pressure Phase |
|---|---|---|---|---|---|
|  | **Expt.** | **SCAN** | **PBEsol** | **PBESol + D3+BJ** | **Expt.** |
| KReO$_4$ | $K_0$=28.8(6) $V_0$=404.1(5) | $K_0$=36.2(15) $V_0$=408.0(1) | $K_0$=29.6(7) $V_0$=414.8(1) | $K_0$=30.4(5) $V_0$=422.0(1) | $K_0$=34.1(8) $V_0$=398.9(9) |
| RbReO$_4$ | $K_0$=19.5(7) $V_0$=451.1(5) | $K_0$=30.5(6) $V_0$=448.0(1) | $K_0$=22.9(2) $V_0$=454.8(1) | $K_0$=24.3(2) $V_0$=464.0(1) | $K_0$=29.4(4) $V_0$=441.1(5) |
| AgReO$_4$ | $K_0$=56.2(9) $V_0$=341.1(5) | $K_0$=68.1(11) $V_0$=328.0(5) | $K_0$=61.9(10) $V_0$=322.6(1) | $K_0$=48.7(14) $V_0$=340.0(1) | $K_0$=56.2(9) $V_0$=341.1(5) |

AgReO$_4$ ($K_0$ = 56.2(9) GPa) has the highest bulk modulus followed by KReO$_4$ ($K_0$ = 28.8(6) GPa) and RbReO$_4$ ($K_0$ = 19.5(7) GPa); this order follows that of the ionic radii of the *A*-site cation. Notice that the bulk modulus inversely correlates to the unit-cell volume which is correlated to the volume of the *A*O$_8$ polyhedron as observed in most *AM*O$_4$ oxides [1, 5, 35, 36]. The unit-cell volume and the *A*O$_8$ volume decreases in going from Rb to K and to Ag while the bulk modulus increases following the Rb, K, Ag sequence. The largest bulk modulus for AgReO$_4$ is supported by the charge density analysis by Mukherjee *et al.* that indicates ionic bonding in AgReO$_4$ [10]. The sequence obtained for the bulk modulus of the three compounds is also consistent with the empirical equation for the bulk modulus proposed by Errandonea and Manjón [1]. This equation depends on the average *A*-O bond distances and the effective valence of the *A*-site cation. The bulk modulus values calculated using the model proposed by Errandonea and Manjon are 38 GPa for AgReO$_4$ > 27 GPa for KReO$_4$ > 23 GPa for RbReO$_4$. These results qualitatively agree with our experimental and computational results. The pressure dependence of the volume of the HP phase of AgReO$_4$ can be described by the same EoS found for the low-pressure phase. The bulk modulus for the monoclinic structures of KReO$_4$ and RbReO$_4$, 34.1(8) and 29.4(4) respectively, are slightly larger than the values obtained for the low-pressure phase, which is consistent with the increased density of the HP phase resulting from the discontinuous volume reduction. These results are also supported by the DFT calculations where the three studied approximations show similar trends (see Table 5). Previous DFT calculations performed for CsReO$_4$, RbReO$_4$, KReO$_4$, and NaReO$_4$ [42] also support our conclusions, with the bulk modulus following the sequence CsReO$_4$ < RbReO$_4$ < KReO$_4$ < NaReO$_4$ as expected from the arguments presented in this section.

### 3.3 Elastic constants

To gain a deeper insight into the mechanical behavior of scheelite-type perrhenates we have computed their elastic constants C$_{ij}$. The tetragonal scheelite structure has seven independent elastic constants C$_{11}$, C$_{12}$, C$_{13}$, C$_{16}$, C$_{33}$, C$_{44}$, and C$_{66}$. The values obtained for the three studied compounds, using PBEsol+D3+BJ, are summarized in Table 6. They satisfy the Born stability criteria for the stability of a tetragonal system [43] : C$_{11}$ - C$_{12}$ > 0; 2 C$_{13}^2$ < C$_{33}$(C$_{11}$ + C$_{12}$); C$_{44}$ > 0; C$_{66}$ > 0; 2C$_{16}^2$ < C$_{66}$(C$_{11}$-C$_{12}$). From the elastic constants we obtained the bulk (B), shear (G), and Young (E) modulus, and the Poisson ratio (ν) using the Hill approximation [44].

**Table 6:** Calculated elastic constants and mechanical moduli of AgReO$_4$, KReO$_4$, and RbReO$_4$ at 0 GPa. All magnitudes are given in GPa except for ν which is dimensionless.

| AgReO$_4$ | | KReO$_4$ | | RbReO$_4$ | |
|---|---|---|---|---|---|
| C$_{11}$ = 62.95 | B = 42.11 | C$_{11}$ = 34.25 | B = 21.28 | C$_{11}$ = 30.56 | B = 17.98 |
| C$_{12}$ = 29.17 | G = 13.40 | C$_{12}$ = 15.92 | G = 11.86 | C$_{12}$ = 17.65 | G = 8.96 |
| C$_{13}$ = 34.73 | E = 36.34 | C$_{13}$ = 14.32 | E = 30.00 | C$_{13}$ = 12.10 | E = 23.06 |
| C$_{33}$ = 55.78 | ν = 0.36 | C$_{33}$ = 33.93 | ν = 0.26 | C$_{33}$ = 21.76 | ν = 0.28 |
| C$_{44}$ = 12.44 | | C$_{44}$ = 13.83 | | C$_{44}$ = 10.64 | |
| C$_{66}$ = 14.21 | | C$_{66}$ = 14.70 | | C$_{66}$ = 12.41 | |
| C$_{16}$ = 0.03 | | C$_{16}$ = -3.55 | | C$_{16}$ = -2.56 | |

In the three compounds we found that C$_{11}$ > C$_{33}$ which means that the lattice is more rigid along the *a*-axis compared to the *c*-axis, aligning with the reduction in the *c*/*a* ratio under compression shown in Fig. 8. The bulk moduli obtained from the elastic constants follows the same trend (AgReO$_4$ > KReO$_4$ > RbReO$_4$) to the values obtained from the pressure dependence of the volume (see Table 4). However, the values obtained from the elastic constants 42.11 GPa for AgReO$_4$, 21.28 GPa for KReO$_4$, and 17.98 GPa for RbReO$_4$ are slightly smaller than the values obtained from the Birch-Murnaghan fit to the corresponding DFT calculations, 48.7 GPa, 30,4 GPa, and 24.3 GPa, respectively, and to the experimentally determined values of 56.2, 28.8 and 19.5 GPa. The Young's modulus for the three compounds is of the same order as the bulk modulus, with B > E in AgReO$_4$ and E > B in the other two compounds. This means that in scheelite-type perrhenates, the tensile or compressive stiffness of the material when subjected to a longitudinal force, is similar to its ability to withstand bulk compression. The value of the shear modulus, which is smaller than E and B in the three compounds, implies that they are more prone to shear deformations than to volume reduction, rendering it highly sensitive to non-hydrostatic stress. This is possibly the origin of the anomalous results obtained by Otto *et al*. [8]). The Poisson ratio of the three compounds indicates that they exhibit ductile characteristics.

## 4. Conclusions

By means of high-pressure powder XRD we found that the three studied perrhenates undergo a phase transition from tetragonal to monoclinic symmetry. In KReO$_4$ and RbReO$_4$ this involves a first-order phase transition from scheelite (*I*4$_1$/$_a$) to M'-fergusonite (space group *P*2$_1$/*c*) and in AgReO$_4$ it occurs by a continuous phase transition to M-fergusonite (space group *I*2/*a*). The reported transitions provide an explanation for changes observed in the Raman spectra reported in previous studies [6, 8]. From the pressure dependence of the volume, a Birch-Murnaghan equation of state [33] was determined, and the anisotropic compressibility of the different phases described. RbReO$_4$ has the smallest value of bulk modulus, and AgReO$_4$ the highest among the three compounds. Density-functional theory calculations were also performed. The calculations correctly describe the pressure dependence of the lattice parameters of the scheelite phase. Information on the elastic constants was also obtained. Despite correctly describing the structures of the low-pressure phase, including the pressure dependence of the unit-cell parameters, DFT could not capture the pressure-driven phase transitions. It is thought that this may be a result of pressure-induced delocalization of the *f*-electrons of Re, a phenomenon usually not capture properly by DFT. Further studies are needed to clarify why DFT can correctly describe the properties of the low-pressure scheelite phase but not capture the phase transition induced by pressure.

**Data Availability**

The data that support the findings of this study are available from the corresponding author upon reasonable request.

**Author Information**

**Corresponding Author**

Daniel Errandonea - Departamento de Física Aplicada, Instituto de Ciencias de Materiales, MALTA Consolider Team, Universitat de Valencia, 46100 Valencia, Spain; https://orcid.org/0000-0003-0189-4221; Email: daniel.errandonea@uv.es

**Authors**

Neha Bura - Departamento de Física Aplicada, Instituto de Ciencias de Materiales, MALTA Consolider Team, Universitat de Valencia, 46100 Valencia, Spain.

Pablo Botella - Departamento de Física Aplicada, Instituto de Ciencias de Materiales, MALTA Consolider Team, Universitat de Valencia, 46100 Valencia, Spain.

Catalin Popescu – CELLS-ALBA Synchrotron Light Facility, Cerdanyola del Vallès, 08290 Barcelona, Spain.

Frederico Alabarse – Frederico G. Alabarse - Elettra Sincrotrone Trieste, Italy.

Ganapathy Vaitheeswaran – School of Physics, University of Hyderabad, Prof. C. R. Rao Road, Gachibowli, Hyderabad 500046, Telangana, India.

Alfonso Muñoz – Departamento de Física, MALTA Consolider Team, Universidad de La Laguna, San Cristóbal de La Laguna, Tenerife E-38200, Spain.

Brendan J Kennedy - School of Chemistry, The University of Sydney, Sydney, New South Wales 2006, Australia.

Jose Luis Rodrigo Ramon – - Departamento de Física Aplicada, Instituto de Ciencias de Materiales, MALTA Consolider Team, Universitat de Valencia, 46100 Valencia, Spain.

Josu Sanchez-Martin – - Departamento de Física Aplicada, Instituto de Ciencias de Materiales, MALTA Consolider Team, Universitat de Valencia, 46100 Valencia, Spain.

**Author Contributions**

N.B. formal analysis, methodology, data acquisition, writing review and editing. P.B., C.P., F.A., J.L.R.R., and J.S.M. methodology, data acquisition, writing review and editing. G.V. and A.M. computational calculations, writing review and editing. B.J.K: methodology, writing review and editing. D.E. conceptualization, formal analysis, writing original draft, funding acquisition, writing original draft, formal analysis, writing review and editing. All authors participated in the writing and editing of the manuscript. All authors have given approval to the final version of the manuscript.

**Notes**

The authors declare no competing financial interest.

**Acknowledgements**

Part of this research is supported by the Spanish Ministry of Science, Innovation, and Universities (MCIN/AEI/10.13039/501100011033) under grant number PID2022-138076NB-C41/44, and by Generalitat Valenciana under grants CIPROM/2021/075 and MFA/2022/007. This study is part of the Advanced Materials program supported with funding from the European Union NextGenerationEU (PRTR-C17.I1). The authors thank Elettra Sincrotrone Trieste (Proposal 20230054) and ALBA synchrotron (Proposal 2024028151) for providing beam time for the HP XRD experiments. The research leading to this result has been co-funded by the project NEPHEWS under Grant Agreement No 101131414 from the EU Framework Programme for Research and Innovation Horizon Europe. D.E. and P.B. thank the financial support of Generalitat Valenciana through grant CIAPOS/2023/406. G. V acknowledges Institute of Eminence, University of Hyderabad (UOH-IOE-RC3-21-046) for the financial support and CMSD, University of Hyderabad for providing the computational facility.